\documentclass[namedreferences,hyperref,optionalrh]{spr-sola}
\usepackage{graphicx}        
\usepackage{color}           

\chardef\us=`\_

\usepackage{mathastext}

\begin{document}

\begin{frontmatter}
\title{SynCOM: A tool for simulating coronal outflows}

\author[addressref={aff1,aff2},email={moraesfilho@cua.edu}]{\inits{V.M.F.}\fnm{Valmir}~\snm{Moraes Filho}\orcid{0000-0002-5447-9964}}
\author[addressref={aff1,aff2},email={uritsky@cua.edu}]{\inits{V.M.U.}\fnm{Vadim}~\snm{Uritsky}\orcid{987-654-3210}}
\author[addressref={aff1}]{\inits{B.T.}\fnm{Barbara}~\lnm{Thompson}}
\author[addressref={aff3}]{\inits{S.G.}\fnm{Sarah}~\lnm{Gibson}}
\author[addressref={aff4}]{\inits{C.D}\fnm{Craig}~\lnm{DeForest}}

\address[id=aff1]{NASA Goddard Space Flight Center, Greenbelt, Maryland, USA }
\address[id=aff2]{Department of Physics, Catholic University of America, Washington, DC, USA}
\address[id=aff3]{University Corporation for Atmospheric Research, Boulder, Colorado, USA}
\address[id=aff4]{Southwest Research Institute, Boulder, Colorado, USA}

\runningauthor{Moraes Filho et al.}
\runningtitle{\textit{Solar Physics} Example Article}

\begin{abstract}
SynCOM is a package of procedures written in IDL (Interactive Data Language) that simulates transient solar wind flows. Each function within SynCOM handles specific tasks, such as initializing parameters, generating synthetic profiles, creating Gaussian blobs to represent solar wind features, and producing high-resolution images of the solar corona. This modular design allows users to call or customize individual functions independently, providing flexibility to adjust simulations to different observational or solar wind conditions. The software architecture is designed to facilitate SynCOM, which effectively creates synthetic datasets for testing and verifying feature tracking algorithms. It also takes advantage of the robust capabilities of the IDL for high-performance scientific computing.
 
\end{abstract}
\end{frontmatter}

\section{Introduction}  
\label{S-introduction}
The challenge of mapping the global flow of the solar wind, especially in the outer corona and inner heliosphere, remains a key issue in heliophysics. The Flow Tracking Challenge has already shown the necessity and effectiveness of a testbed using the initial SynCOM model to refine the tracking algorithms. This highlights the need for further development of robust models for algorithm validation.

While the upcoming PUNCH mission aims to capture solar wind structures with unprecedented resolution, a testbed like SynCOM is crucial for providing a ground truth for flow-tracking methods. Tracking solar wind features is complex because of its variable nature, which causes uncertainties in velocity measurements. The Flow Tracking Challenge revealed that feature-tracking algorithms, such as optical flow and distance-time (DT) plotting, lack accuracy without a ground-truth reference. SynCOM, Synthetic Corona Outflow Model, addresses this need with a data-driven approach that bridges observational data with solar wind dynamics.




\section{SynCOM: The algorithm}
\label{S-SynCOM}

SynCOM is essential for mapping solar wind outflows by integrating simulations with real observations. This framework allows for precise comparisons between feature tracking algorithms, ensuring that future data from missions like PUNCH can be effectively used to understand and map solar wind dynamics.

SynCOM is a data-driven statistical model designed to emulate the key attributes of transient solar wind flows without relying on complex physics-based equations. Its core principle is to provide a fast and memory-efficient solution capable of generating numerous simulations of the dynamic solar corona. These simulations, configured with specific kinematic parameters, help facilitate the interpretation of observational data and the validation of flow tracking methodologies in the scientific community.

A central goal of SynCOM is to create a synthetic dataset for training and testing solar wind tracking algorithms. Current algorithms, developed on observational data, lack a "ground truth" to verify their accuracy. SynCOM addresses this limitation by providing synthetic data with spatial and temporal properties statistically similar to those of real observations, but with known local flow speeds. This allows developers to evaluate the performance of their algorithms, identifying areas where improvements are needed based on the ground truth provided by SynCOM.

The IDL source code for the SynCOM software suite is available at zenodo.org/records/13357546 \citep{moraes_filho_2024}, offering access to the high-resolution simulations of transient solar wind flows described in this work.

\subsection{The Propagating Gaussian blobs} 
    \label{S-Description}

The design of SynCOM was inspired by simulating the solar wind outflow emanating from the Sun. This led to the representation of solar wind clumps as radially expanding two-dimensional Gaussian perturbations. In this study, these structures are based on observations of plasma blobs that move outward from the corona, as documented by \cite{Sheeley_2009}. According to their findings, these irregularities begin around 3-4 $R_\odot$ from the center of the Sun as compact blobs of material, approximately 1 $R_\odot$ in length and 0.1 $R_\odot$ in width, separating from the tips of coronal streamers. The propagating Gaussian blob is a mathematical model of a transient propagating coronal feature represented in plane polar coordinates \citep{moraes_filho_2024}:
\begin{equation}
\label{eq:gaussian_blob}
    G(\theta, r) = G_0 \exp{\left(-\left[{\frac{(\theta-\theta_0)^2}{2L_\theta^2}}+{\frac{(r - r_0(t))^2}{2L_r^2}}\right]\right)} ,    
\end{equation}
where $\theta$ and $r$ are respectively the position angle ($\theta$, measured counterclockwise from the North pole at the solar limb) and the radial coordinates of the Gaussian blob, $\theta_0$ is the central $\theta$ position of the blob,  $r_0(t)$ is its central radial position evolving as a function of time $t$, $L_\theta$ and $L_r$ are the characteristic sizes of the blob in the $\theta$ and $r$ direction, correspondingly, and $G_0$ is the peak intensity of the blob. For the purpose of this paper, $\theta_0$ is chosen to be time independent, which results in the strictly radial propagation of the blobs.

As defined in equation \ref{eq:gaussian_blob}, each Gaussian blob is a two-dimensional intensity array that changes over time and is determined by a four-element vector of adjustable parameters $[\theta_0, r_0, L_\theta, L_r]$, which specify the blob's position and shape.

\begin{figure}
    \centering
    \includegraphics[width=\linewidth]{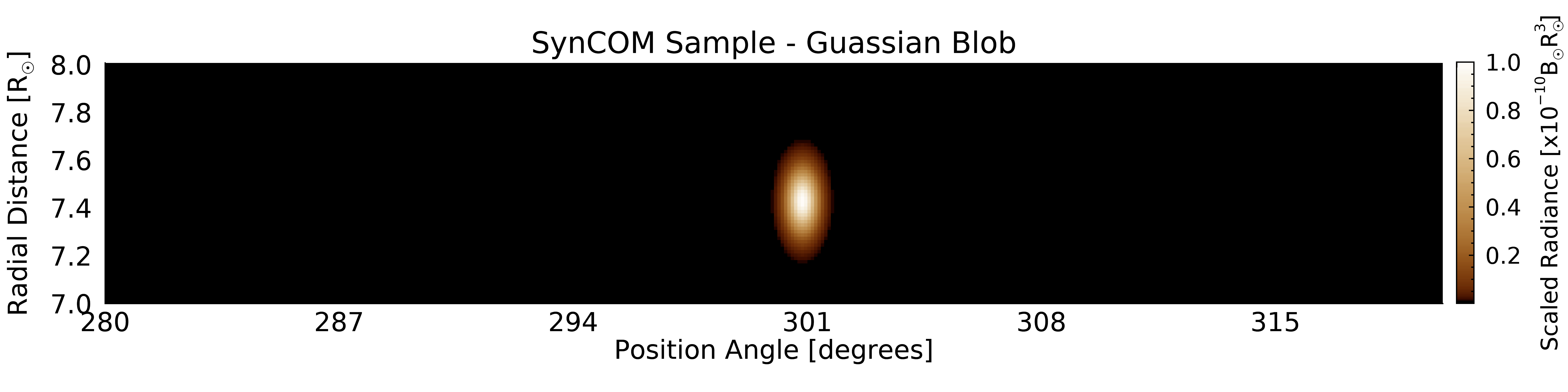}
    \caption{SynCOM blob example. This representation displays a Guassian blob sample, designed to emphasize its shape and size. In this instance, the brighter blob stands out in the image, appearing to be centered at angle 301 degrees and extending from 7.2 to 7.6 solar radii.}
    \label{fig:simple_sample}
\end{figure}

Figure \ref{fig:simple_sample} shows a Gaussian blob, originating from $\theta=301^\circ$ with a core at $7.4 R_\odot$. This paper uses the Heliocentric Radial Coordinate System as described by \cite{Thompson_2006}. The subsequent discussion will elucidate the methodology used by the model to generate a single simulated image.

We have developed a method to produce synthetic high-resolution images of solar wind density by simulating the motion of multiple Gaussian blobs within the solar corona. 

Having established the SynCOM theoretical framework, we now discuss its computational implementation in IDL modules. These modules sequentially handle parameter initialization, data loading, blob creation, and synthetic image generation, all aimed at simulating solar wind outflows. The following discussion will elaborate on the function of each module in this process.

\section{Modules}

This section outlines the key components of the SynCOM model, which are responsible for simulating solar wind dynamics. Each module serves a distinct purpose, from initializing essential parameters to generating high-resolution synthetic images. This section details the functionality of each module, including parameter setup, data loading, simulation core processes, and the generation of the final visual output. These modules work together to create an efficient framework that can simulate solar wind flows, providing researchers with a powerful tool for studying transient solar phenomena.

\subsection{Parameter Initialization}

Calling sequence: \texttt{SYNCOM\_PRAMS,ModPramsStruct,SYNCOM\_N\_BLOBS = $n_{blobs}$,}\\
\texttt{SYNCOM\_NX = $n_\theta$, SYNCOM\_NY = $n_r$,}\\
\texttt{SYNCOM\_CADENCE = $300.$, SYNCOM\_PIXEL = $9744.$}
\newline
\newline
The \texttt{SYNCOM\_PRAMS} module is responsible for initializing the essential parameters required to run the SynCOM simulation. These are the key parameters required to start its operation, such as the number of blobs ($n_{blobs}$) released, the number of pixels in each dimension of the image ($n_\theta, n_r$), and the initial launch position of each blob ($r_0$), all of which are stored in the \texttt{ModPramsStruct}. In the example calling sequence, the image dimensions default to the resolution of COR2 images.

Additionally, the module allows for the inclusion of optional parameters that provide greater control over the physical properties of the simulation. They include the time cadence and the pixel size derived from specific instruments, in particular the COR2 instrument, which operates with a $300$ second cadence and a pixel size of $9744$ kilometers. Other parameters, such as acceleration, peak brightness intensity, and noise level, enable more precise adjustments to the simulation environment.

The values of these parameters are either obtained from observational data or estimated using datasets such as those from the COR2 instrument \citep{Howard_2008} on board the STEREO spacecraft \citep{Kaiser_2008}.

\subsection{Loading Profiles}

Calling sequence: \texttt{SYNCOMLOAD, ModPramsStruct, LoadStruc}
\newline
\newline
The \texttt{SYNCOMLOAD} module is responsible for loading the profiles necessary for the simulation, such as the period, radius, and velocity of the central position of each blob ($\theta_0$). These profiles can be derived either from observational data or imported from external sources, allowing flexibility in how the model is used. This versatility enables SynCOM to be easily adapted to various types of datasets and research needs.

The period, $T$, determines the frequency of blob appearances, where each blob surfaces once during each period. The radius, $L$, dictates the size of the blobs, while the velocity, $V(\theta)$, affects their central position as time, $t$, progresses. For example, currently, the model uses a predefined function to determine the central location of the blobs for their initial launch,
\begin{equation}
\label{eq:blob_final_position}
    r_0(t)=r_B + V(\theta) t
\end{equation}
where $r_B$ is the blob's position at the observation boundary, $5 R_\odot$, and $V(\theta)$ is the radial velocity for a specific $\theta_0$. Future versions of SynCOM aim to incorporate radial acceleration, which requires time-domain integration of Equation \ref{eq:blob_final_position}, making the model more dynamic.

Each blob is then assigned with a $\theta_0$, which obtains its own set of variables from the database, all of which are stored in the \texttt{LoadStruc}. This data-driven approach ensures that each blob has distinct values for velocity, period, radius, and initial radial and angular positions, allowing for the creation of a diverse and realistic simulation of solar wind structures.

\subsection{Simulation Core}

Calling sequence: \texttt{SYNCOM\_CORE, N\_BLOBS = N\_BLOBS, syncom\_data,}\\
\texttt{time\_t = $t$, time0 = $t_0$, syncom\_version, scale\_factor = scale}
\newline
\newline
The \texttt{SYNCOM\_CORE} module is the core of the simulation, which coordinates the integration of all modules into a cohesive process. The simulation output is stored in \texttt{syncom\_data}, which represents the data cube that captures the time-dependent evolution of the solar wind structures. Before creating a blob, it is necessary to convert the velocity, radial position, and radius into pixels. This conversion is essential to accurately position the blobs in the simulation space, allowing the model to replicate their real-world movement through the solar corona.

The \texttt{syncom\_version} parameter specifies the version name for the simulation output, which can be customized by the user based on their needs. The simulation begins at \texttt{time0} (the initial time of the simulation) and runs until \texttt{time\_t} (the final time of the simulation), allowing users to control the duration of the simulated period. Additionally, the \texttt{scale\_factor} parameter adjusts the size of the blobs, offering flexibility in the granularity of the simulation, from coarse-grained to fine-scale simulations.

The resulting pixel-based parameters enable precise tracking of each blob's evolution in the grid over time, ensuring that the simulation accurately captures the solar wind's transient behavior.

\subsection{Image Generation} 

Calling sequence: \texttt{SYNCOMIMAGE, ModPramsStruct, syncom\_version,}\\ 
\texttt{time\_0, radial\_i, v\_array, period\_array, L\_array, PSI, time\_t, img}
\newline
\newline
The \texttt{SYNCOMIMAGE} module processes each blob individually using the specific variables assigned to it by the \texttt{SYNCOMLOAD} module. The creation of the blob is based on equation \ref{eq:gaussian_blob}, which defines the structure of the blob. These inputs include \texttt{radial\_i} which is the initial radial position of the blob, controlling its central position in the $r$-direction, \texttt{v\_array} the velocity for each position angle, which determines the movement of the blob's radial center over time according to equation \ref{eq:blob_final_position}, \texttt{period\_array} defines how often the blob reappears in time, \texttt{L\_array} controls the blob size, and \texttt{PSI} is the position angle array, which controls the blob's central position in the $\theta$-direction. Each blob is periodically created and shifted radially to simulate its movement through the solar corona, generating multiple instances at regular intervals to reflect the quasiperiodic nature of the solar wind.

Though \texttt{SYNCOMIMAGE} is not directly invoked in the main process, it is repeatedly called by \texttt{SYNCOM\_CORE} at each time step, progressing from \texttt{time\_0} to \texttt{time\_t}. The module accumulates the density distributions of all blobs in an initially empty image array (\texttt{img}), which builds over time as more blobs are added.

This process continues until all blobs are processed, producing an image that captures the complexity of solar wind flows. The final image, which includes both small- and large-scale dynamics, is saved as an FITS file. The \texttt{syncom\_version} input determines the file name, allowing users to label the output according to the simulation version or specific identifiers.

The \texttt{scale\_factor} input, provided by the \texttt{SYNCOM\_CORE} procedure, modifies \texttt{L\_array}, adjusting the blob size and enabling the simulation to capture either small-scale or large-scale dynamics (fine-scale or coarse-grained). This flexibility allows researchers to tailor the simulation resolution to the needs of their study. The FITS file format ensures high-resolution preservation for future analysis, offering a detailed set of synthetic images to study the transient dynamics of solar wind structures.

\subsection{Visual Fidelity}

Calling Sequence: \texttt{SYNCOMNOISE, ModPramsStruct, img, new\_img, }\\
\texttt{file\_name, /noise\_add, /luminosity\_add,}\\
\texttt{rect\_img, /rectangular}
\newline
\newline
The \texttt{SYNCOMNOISE} module is designed to enhance the visual accuracy of the synthetic solar wind images by applying three key processes: noise addition, brightness variation, and coordinate transformation. Each of these processes is activated independently using specific keywords, allowing users to fine-tune the output image to match observational data.

If the \texttt{noise\_add} keyword is set, random noise is added to the image. This simulates the instrument and random noise that is typically encountered in real observational data. The noise level is determined by the value provided in the \texttt{ModPramsStruct}, ensuring consistency with the model parameters. This function is essential for producing images that resemble those captured by the instruments, adding realism to the simulation.

If the \texttt{luminosity\_add} keyword is set, the module modifies the brightness behavior of the image. This process introduces intensity variations by applying an inverse normalization procedure based on predefined intensity profiles derived from the COR2 data. The image pixels are adjusted according to their mean and standard deviation values, replicating the luminosity patterns observed in real data. This allows the synthetic images to capture realistic brightness fluctuations in solar wind structures.

If the \texttt{rectangular} keyword is set, the module converts the image from polar to rectangular coordinates. This transformation is useful for cases where a rectangular grid is required for further analysis or visualization. The function uses a transformation to create a rectangular representation of the solar wind images, maintaining the spatial relationships present in the polar data.

Each of these processes can operate independently or in combination, making the \texttt{SYNCOMNOISE} module flexible and adaptable to different visualization and analysis needs. The modified image is saved in \texttt{new\_img}, and if the rectangular transformation is applied, the transformed image is stored in \texttt{rect\_img}. Optional parameters such as \texttt{file\_name} can be provided for the intensity profile statistics. This module allows users to apply a variety of visual modifications to the simulation, helping to simulate realistic observational effects for synthetic solar wind images.

\section{Functions}

This section details the functions included in the SynCOM package. Each function has a distinct role, such as generating propagating blobs and converting images from polar to Cartesian coordinates. These functions are executed within specific modules whenever they are called, specifically within the modules \texttt{SYNCOMIMAGE} and \texttt{SYNCOMNOISE}.

\subsection{Gaussian Blob}

Calling Sequence: \texttt{result = GAUSSIANWAVE(npixel=npixel,}\\
\texttt{avr=avr, st\_dev=st\_dev)}
\newline
\newline
The \texttt{GAUSSIANWAVE} function generates a 1D or 2D Gaussian wave to simulate solar wind features propagating through space. The Gaussian is defined by the pixel dimensions (\texttt{npixel}), central position (\texttt{avr}), and standard deviation or width (\texttt{st\_dev}). The SYNCOM framework relies heavily on this function, as it is called each time \texttt{SYNCOMIMAGE} produces a new synthetic image for simulations.

For a 1D Gaussian, \texttt{npixel} specifies the number of pixels along the x-axis as [\texttt{nx}]. The central position and standard deviation are given as [\texttt{avr\_x}] and [\texttt{st\_dev\_x}], respectively. The function then generates a 1D Gaussian curve based on these values.

For a 2D Gaussian, \texttt{npixel} is specified as [\texttt{nx}, \texttt{ny}], where \texttt{nx} and \texttt{ny} represent the number of pixels along the x and y axes, respectively. The central positions and standard deviations are provided as [\texttt{avr\_x}, \texttt{avr\_y}] and [\texttt{st\_dev\_x}, \texttt{st\_dev\_y}], respectively. The function generates a 2D Gaussian surface based on these inputs.

Either a 1D or 2D array is generated for the resulting Gaussian wave based on the input. This function mainly serves to simulate propagating blobs in solar wind studies, with the Gaussian wave depicting a smooth, symmetrical feature in motion over time. In the case of 2D waves, the position of the blob moves along the y axis, enabling the modeling of dynamic and time-evolving solar wind phenomena in the radial direction.

\subsection{Coordinate Transformation}

Calling Sequence: \texttt{result = POLAR\_TO\_RECT\_TR(img\_p,}\\
\texttt{phi\_range, rho\_range, n=n)} 
\newline
\newline
The \texttt{POLAR\_TO\_RECT\_TR} function transforms a polar image into a rectangular (Cartesian) coordinate system via triangulation. In contrast to \texttt{GAUSSIANWAVE}, which is consistently called by \texttt{SYNCOMIMAGE}, \texttt{POLAR\_TO\_RECT\_TR} is only executed by \texttt{SYNCOMNOISE} when the \texttt{rectangular} keyword is enabled. Once enabled, this function creates an image format that is completely different from the one generated by the \texttt{SYNCOM\_CORE} workflow.

The input is a 2D array, \texttt{img\_p}, where the first dimension represents azimuthal angles and the second corresponds to radial distances. The transformation relies on two key input parameters: \texttt{phi\_range}, defining the minimum and maximum azimuthal angles (in degrees) and \texttt{rho\_range}, specifying the minimum and maximum radial distances (e.g. solar radii). An optional parameter n allows users to control the number of bins in the x and y directions of the output rectangular image, with a default value of 1024.

To perform the conversion, the function translates each pixel's azimuthal and radial coordinates into Cartesian coordinates. These coordinates are then used to triangulate and interpolate the pixel values onto a rectangular grid. A mask is applied to the region within the minimum radial distance, where the pixel values are set to zero to simulate the central mask of the Sun, ensuring that the output accurately reflects the structure of the original image.

The function ultimately returns the transformed rectangular image, the result, which can be further analyzed or visualized in a Cartesian framework. This polar-to-rectangular conversion is crucial for facilitating a deeper analysis of solar wind features in Cartesian coordinates, enabling enhanced interpretation and exploration of the data.

\section{Example Workflow}
\label{S-workflow}

\begin{figure}
    \centering
    \includegraphics[width=\linewidth]{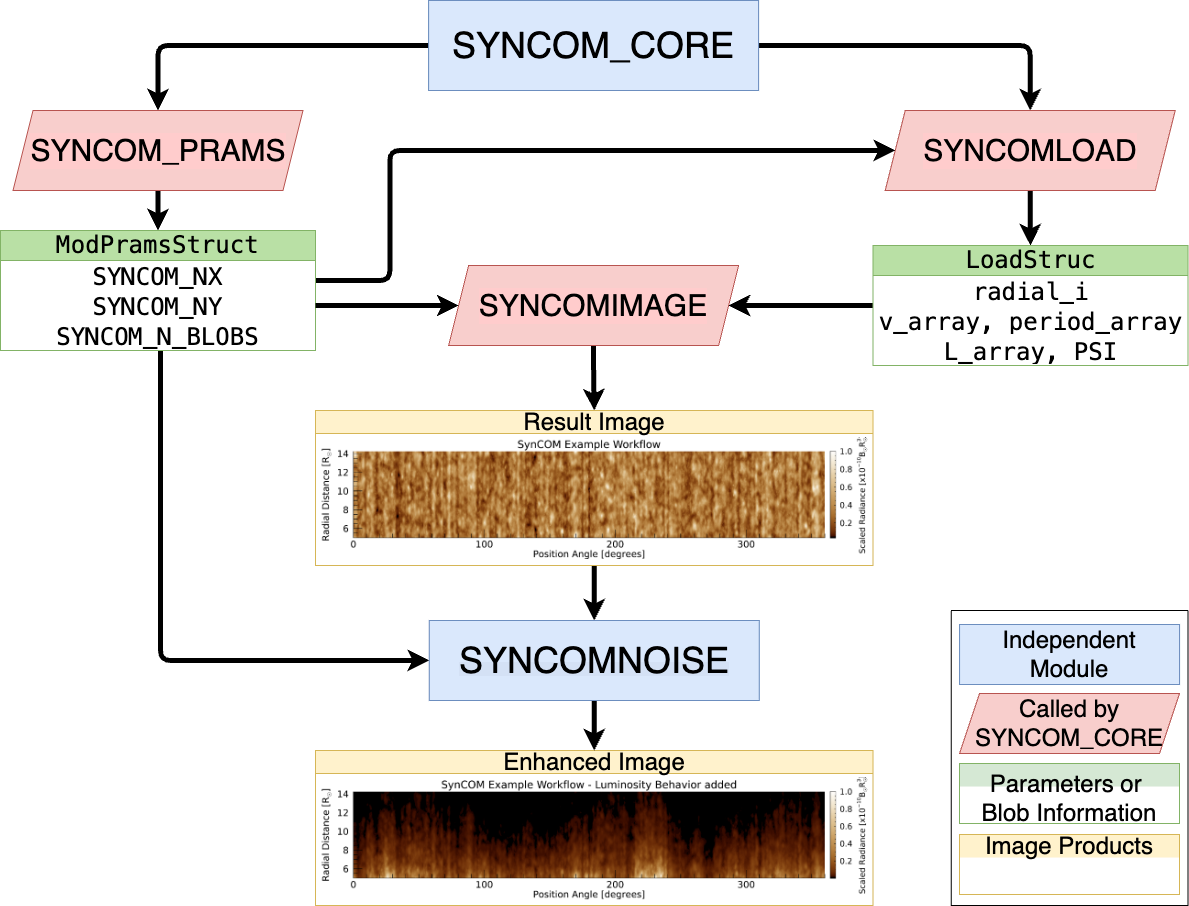}
    \caption{Flowchart illustrating the workflow of the SynCOM framework for generating and enhancing synthetic solar wind images. \texttt{SYNCOM\_CORE} orchestrates the process by integrating two key structures: \texttt{ModPramsStruct} and \texttt{LoadStruc}. First, \texttt{SYNCOM\_CORE} initializes parameters stored in \texttt{ModPramsStruct}, which is handled internally by \texttt{SYNCOM\_PRAMS}. Next, blob properties such as positions, velocities, and sizes are loaded from \texttt{LoadStruc} via \texttt{SYNCOMLOAD}. After converting these profiles into pixel units, \texttt{SYNCOM\_CORE} calls \texttt{SYNCOMIMAGE} to generate the synthetic solar wind images for each time step, which are stored in the \texttt{syncom\_data} array, resulting in the "Result Image." Finally, \texttt{SYNCOMNOISE} is applied independently to enhance the visual fidelity of the images by adding brightness adjustments, producing the "Enhanced Image". This modular workflow allows flexible image generation and post-processing.}
    \label{fig-syncom-workflow}
\end{figure}

To generate a set of synthetic solar wind images and enhance them with brightness adjustments, as shown in Figure \ref{fig-syncom-workflow}, follow this two-step process:
\begin{itemize}
    \item[1.] \textbf{Generate Synthetic Images}:\\
    The \texttt{SYNCOM\_CORE} procedure is responsible for generating synthetic solar wind images. It can handle the initialization of parameters and load the necessary structures internally. To use it, provide the number of blobs, the time frame, and optional parameters such as version and scale factor. For example: 
\newline 
    \texttt{SYNCOM\_CORE,N\_BLOBS = $1000.$,syncom\_data,time\_t = $100$,time0 = $0$, }\\
    \texttt{syncom\_version = "example", scale\_factor = $1.0$}
\newline  
    This command will create a set of synthetic solar wind images, saving them with the prefix "example". The 3D array \texttt{syncom\_data} will store the simulated images as well as the parameters, including \texttt{ModPramsStruct} and \texttt{LoadStruc}.
\newline  
    \item[2.] \textbf{Enhance the Simulated Images}: \\
    Once the synthetic images are generated, the \texttt{SYNCOMNOISE} procedure can be used to add noise and simulate brightness variations based on predefined intensity profiles. This step mimics real observational conditions and improves the visual fidelity of the images. For example:
\newline 
    \texttt{SYNCOMNOISE,ModPramsStruct,syncom\_data,new\_img,/luminosity\_add}
\newline 
    n this case, the enhanced image will be stored in \texttt{new\_img}, with added brightness adjustments because the \texttt{/luminosity\_add} keyword was set.
\end{itemize}

\section{Conclusion} 
\label{S-Conclusion} 

The development of SynCOM was aimed at addressing the complex task of simulating transient solar wind flows, particularly the challenge of capturing diverse solar wind phenomena with a robust and adaptable framework. SynCOM's design is built around Gaussian blobs, which represent dynamic solar wind features. This simple yet effective structure allows for a broad range of solar wind behaviors, from small-scale jetlets to large-scale coronal mass ejections (CMEs).

The key strengths of SynCOM include its open source architecture, which enables users to modify model parameters and customize simulations. This adaptability facilitates a diverse range of research applications. For instance, custom simulations conducted with user-defined parameters showed the model's versatility in accurately reproducing various solar wind features. These custom runs validated the model's ability to simulate transient solar wind structures across different scales and velocities.

In conclusion, customized simulations demonstrated that SynCOM is an invaluable tool for solar wind research, providing a reliable and flexible platform for modeling solar wind dynamics. These findings underscore SynCOM's critical role in future solar research, especially with missions like PUNCH, where understanding the small- and large-scale dynamics of the solar wind is essential.

\bibliographystyle{spr-mp-sola}
\bibliography{sola_bibliography_example}

\end{document}